\documentclass[12pt,a4paper]{article}

  \usepackage{a4wide}
  \usepackage{latexsym}
  \usepackage{epsf}
  \usepackage{amssymb}
  \usepackage{graphicx}
  \usepackage{amsmath, cite}
  \usepackage{slashed,epsfig}
  \usepackage{bbm}
  \usepackage{bbm}
  \usepackage{amsmath,amssymb,amsthm}
  \usepackage{pst-all}
  \usepackage{here}

\renewcommand{\d}{\textrm{d}}
\newcommand{\Real}{\textrm{I\!R}}

\newcommand{\e}{\textrm{e}}
\renewcommand{\d}{\textrm{d}}

\newcommand{\SO}{\mathop{\rm SO}}
\newcommand{\CSO}{\mathop{\rm CSO}}

\newcommand{\SL}{\mathop{\rm SL}}

\begin{document}

\begin{flushright}
\small
UG-07-01\\
\date \\
\normalsize
\end{flushright}

\begin{center}

\vspace{.7cm}

{\LARGE {\bf Scaling cosmologies,
\\ \vspace{.4cm}
 geodesic motion and pseudo-susy }} \\

\vspace{1.2cm}

  {\large Wissam Chemissany, Andr\'e Ploegh and Thomas Van Riet} \\[3mm]
{\small\slshape
   Centre for Theoretical Physics, University of Groningen,\\
    Nijenborgh 4, 9747 AG Groningen, The Netherlands} \\
{\upshape\ttfamily w.chemissany, a.r.ploegh, t.van.riet@rug.nl}\\[3mm]

\vspace{5mm}

{\bf Abstract}
\end{center}
One-parameter solutions in supergravity carried by scalars and a
metric trace out curves on the scalar manifold. In ungauged
supergravity these curves describe a geodesic motion. It is known
that a geodesic motion sometimes occurs in the presence of a scalar
potential and for time-dependent solutions this can happen for
scaling cosmologies. This note contains a further study of such
solutions in the context of pseudo-supersymmetry for multi-field
systems whose first-order equations we derive using a
Bogomol'nyi-like method. In particular we show that scaling
solutions that are pseudo-BPS must describe geodesic curves.
Furthermore, we clarify how to solve the geodesic equations of
motion when the scalar manifold is a maximally non-compact coset
such as occurs in maximal supergravity. This relies upon a
parametrization of the coset in the Borel gauge. We then illustrate
this with the cosmological solutions of higher-dimensional gravity
compactified on a $n$-torus.

\begin{quotation}

\small
\end{quotation}

\newpage

\pagestyle{plain} \tableofcontents

\section{Preliminaries}
We consider scalar fields $\Phi^i$ that parametrize a Riemannian
manifold with metric $G_{ij}$ coupled to gravity through the
standard action
\begin{equation}
S=\int
\sqrt{-g}\Bigl\{\mathcal{R}-\tfrac{1}{2}G_{ij}g^{\mu\nu}\partial_{\mu}
\Phi^i \partial_{\nu} \Phi^j -V(\Phi) \Bigr\}\,.
\end{equation}
We restrict to solutions with the following $D$-dimensional
space-time metric
\begin{equation}\label{ansatz}
\d s_D^2= g(y)^2\d s^2_{D-1}+\epsilon f(y)^2\d y^2\,,\quad \d
s^2_{D-1}=(\eta_{\epsilon})_{ab}\d x^a\d x^b\,,
\end{equation}
where $\epsilon=\pm 1$ and
$\eta_{\epsilon}=\text{diag}(-\epsilon,1,\ldots,1)$. The case
$\epsilon= -1$ describes a flat FLRW-space-time and $\epsilon= +1$ a
Minkowski-sliced domain wall (DW) space-time. The scalar fields that
source these space-times can only depend on the $y$-coordinate
$\Phi^i=\Phi^i(y)$. The function $f$ corresponds to the gauge
freedom of reparameterizing the $y$-coordinate.

Of particular interest in this note are scaling comologies, which
have received a great deal of attention in the dark-energy
literature, see \cite{Copeland:2006wr} for a review and references.
One definition (amongst many) of scaling cosmologies is that they
are solutions for which all terms in the Friedmann equation have the
same time dependence. For pure scalar cosmologies this implies that
\begin{equation}
H^2 \sim V \sim T \sim \tau^{-2}\,,
\end{equation}
where $\tau$ denotes cosmic time, $H$ the Hubble parameter and $T$
is the kinetic energy $T=\tfrac{1}{2}\,
G_{ij}\dot{\Phi}^i\dot{\Phi}^j$. These relations imply that the
scale factor is power-law $a(\tau)\sim \tau^p$. In the case of
curved FLRW-universes we also demand that $H \sim k/a^2$, which is
only possible for $p=1$. Interestingly, scaling solutions correspond
to the FLRW-geometries that possess a time-like conformal
vectorfield $\xi$ coming from the transformation
\begin{equation}\label{homothetic}
\tau\rightarrow \e^{\lambda}\tau\,,\qquad x^a\rightarrow
\e^{(1-p)\lambda}x^a\,,
\end{equation}
where $x^a$ are the space-like cartesian coordinates\footnote{For
curved FLRW-space-times the space-like coordinates are
invariant.}. In the forthcoming we reserve the indices
$a,b,\ldots$ to denote space-like coordinates when we consider
cosmological space-times. Apart from the intriguing cosmological
properties of scaling solutions they are also interesting for
understanding the dynamics of a general cosmological solution
since scaling solutions are often critical points of an autonomous
system of differential equations and therefore correspond to
attractors, repellors or saddle points \cite{Copeland:1997et}.
Scaling cosmologies often appear in supergravity theories (see for
instance \cite{Rosseel:2006fs,deRoo:2006ms}) but, remarkably, they
also appear by spatially averaging inhomogeneous cosmologies in
classical general relativity \cite{Buchert:2006ya}.

We will use two coordinate frames to describe scaling comologies
\begin{align}
 \tau-\text{frame}:\qquad& \d s^2=-\d \tau^2 + \tau^{2p}\,\d s^2_{D-1}\,,\\
 t-\text{frame}:\qquad& \d s^2=-\e^{2t}\,\d t^2 + \e^{2pt}\d
 s^2_{D-1}\,.
\end{align}
The first is the usual FLRW-coordinate system and the second can
be obtained by the substitution $t=\ln \tau$.

\section{(Pseudo-) supersymmetry}

If the scalar potential $V(\Phi)$ can be written in terms of another
function $W(\Phi)$ as follows
\begin{equation}\label{susy}
V=\epsilon\,\Bigl\{\tfrac{1}{2}G^{ij}\partial_i W\partial_j W -
\tfrac{D-1}{4(D-2)}W^2\Bigr\} \,,
\end{equation}
then the action can be written as ``a sum of squares'' plus a
boundary term when reduced to one dimension:
\begin{align} S=&\epsilon\int\d y\,
fg^{D-1}\Bigl\{\tfrac{(D-1)}{4(D-2)}\Bigl[W-2(D-2)\frac{\dot{g}}{fg}\Bigr]^2
-\tfrac{1}{2}||\frac{\dot{\Phi}^i}{f} + G^{ij}\partial_j W||^2 \Bigr\} \nonumber\\
& + \epsilon\int
\d\Bigl\{g^{D-1}W-2(D-1)\dot{g}g^{D-2}f^{-1}\Bigr\}\,,
\end{align}
where a dot denotes a derivative w.r.t. $y$. The term
$||\dot{\Phi}^i/f + G^{ij}\partial_j W||^2$ is a shorthand
notation and the square involves a contraction with the field
metric $G_{ij}$. It is clear that the action is stationary under
variations if the terms within brackets are zero\footnote{For
completeness we should have added the Gibbons-Hawking term
\cite{Gibbons:1976ue} in the action which deletes that part of the
above boundary term that contains $\dot{g}$.}, leading to the
following \emph{first-order} equations of motion
\begin{equation}
\boxed{W=2(D-2)\frac{\dot{g}}{fg}\,,\qquad \frac{\dot{\Phi}^i}{f} +
G^{ij}\partial_j W=0\,.}
\end{equation}
For $\epsilon=+1$ these equations are the standard
Bogomol'nyi-Prasad-Sommerfield (BPS) equations for domain walls
that arise from demanding the susy-variation of the fermions to
vanish, which guarantees that the DW preserves a fraction of the
total supersymmetry of the theory. The function $W$ is then the
superpotential that appears in the susy-variation rules and
equation (\ref{susy}) with $\epsilon=+1$ is natural for
supergravity theories. It is clear that for every $W$ that obeys
(\ref{susy}) we can find a corresponding DW-solution, and if $W$
is not related to the susy-variations we call the solutions fake
supersymmetric \cite{Freedman:2003ax}.

For $\epsilon=-1$ these equations are the generalization to
arbitrary space-time dimension $D$ and field metric $G_{ij}$ of
the framework found in references
\cite{Bazeia:2005tj,Liddle:2000cg,Skenderis:2006fb,Skenderis:2006jq}.
So here we generalized and derived in a different way (some of)
the results of
\cite{Bazeia:2005tj,Liddle:2000cg,Skenderis:2006fb,Skenderis:2006jq}
by showing that analogously to DW's we can write the Lagrangian as
a sum of squares. We refer to these first-order equations as
pseudo-BPS equations and $W$ is named the pseudo-superpotential
because of the immediate analogy with BPS domain walls in
supergravity \cite{Skenderis:2006jq, Skenderis:2006fb}. For the
case of cosmologies there is no natural choice for $W$ as
cosmologies cannot be found by demanding vanishing susy-variations
of the fermions\footnote{Star supergravity is an exception
\cite{Hull:2001ii} and that seems related to pseudo-supersymmetry
\cite{Bergshoeff:2007cg}.}.

In \cite{Skenderis:2006jq} it is proven that for all single-scalar
cosmologies (and domain walls) a pseudo-superpotential $W$ exists
such that the cosmology is pseudo-BPS and that one can give a
fermionic interpretation of the pseudo-BPS flow in terms of
so-called pseudo-Killing spinors. This does not necessarily carry
over to multi-scalar solutions as was shown in \cite{Sonner:2007cp}.
Nonetheless, a multi-field solution can locally be seen as a
single-field solution \cite{Celi:2004st} because locally we can
redefine the scalar coordinates such that the curve $\Phi(y)$ is
aligned with a scalar axis and all other scalars are constant on
this solution. A necessary condition for the single-field pseudo-BPS
flow to carry over (locally) to the multi-field system is that the
truncation down to a single scalar is consistent (this means that
apart from the solution one can put the other scalars always to
zero)\cite{Sonner:2007cp}.

\section{Multi-field scaling cosmologies}

Let us turn to scaling solutions in the framework of
pseudo-supersymmetry and see how geodesic motion arises. First we
consider the rather trivial case with vanishing scalar potential $V$
and then in section 3.2 we add a scalar potential $V$.
Pseudo-supersymmetry is only discussed in the case of non-vanishing
$V$.

\subsection{Pure kinetic solutions}
If there is no scalar potential the solutions trace out geodesics
since after a change of coordinates $y\rightarrow \tilde{y}(y)$ via
$\d\tilde{y}=f g^{1-D}\d y$, the scalar field action becomes $\int
G_{ij}\Phi'^i\Phi'^j\d \tilde{y}$, where a prime means a derivative
w.r.t. $\tilde{y}$. This new action describes geodesic curves with
affine parameter $\tilde{y}$. The affine velocity is constant by
definition and positive since the metric is positive definite
\begin{equation}
G_{ij}\Phi'^i\Phi'^j=||v||^2\,.
\end{equation}
The Einstein equation is
\begin{equation}
\mathcal{R}_{yy}=\tfrac{1}{2}G_{ij}\dot{\Phi}^i\dot{\Phi}^j=\frac{||v||^2}{2}g^{2-2D}f^{2}\,,\qquad
\mathcal{R}_{ab}=0\,.
\end{equation}
In the gauge $f=1$ the solution is given by
$g=\e^{C_2}(y+C_1)^{\tfrac{1}{D-1}}$, with $C_1$ and $C_2$ arbitrary
integration constants, but with a shift of $y$ we can always put
$C_1=0$ and $C_2$ can always be put to zero by re-scaling the
space-like coordinates. In the case of a four-dimensional cosmology
the geometry is a power-law FLRW-solution with $p=1/3$.

\subsection{Potential-kinetic scaling solutions}

In a recent paper of Tolley and Wesley an interesting interpretation
was given to scaling solutions \cite{Tolley:2007nq}, which we repeat
here. The finite transformation (\ref{homothetic}) leaves the
equations of motion invariant if the action $S$ scales with a
constant factor, which is exactly what happens for scaling solutions
since all terms in the Lagrangian scale like $\tau^{-2}$. Under
(\ref{homothetic}) the metric scales like $\e^{2\lambda}g_{\mu\nu}$
and in order for the action to scale as a whole we must have
\begin{equation}\label{transfor}
V\rightarrow \e^{-2\lambda}V\,,\qquad
T=\tfrac{1}{2}g^{\tau\tau}G_{ij}\dot{\Phi}^i\dot{\Phi}^j\rightarrow
\e^{-2\lambda}T\,.
\end{equation}
Equations (\ref{transfor}) imply that
$G_{ij}\dot{\Phi}^i\dot{\Phi}^j$ remains invariant from which one
deduces that $\frac{\d \Phi^i}{\d \lambda}=\xi^i$ must be a Killing
vector. The curve that describes a scaling solution follows an
isometry of the scalar manifold. It depends on the parametrization
whether the tangent vector $\dot{\Phi}$ itself is Killing. This
happens for the parametrization in terms of $t=\ln\tau$ since
\begin{equation}
\xi^i = \frac{\d \Phi^i}{\d \lambda} =
\text{lim}_{\lambda\rightarrow 0}
\frac{\Phi^i(e^{\lambda}\tau)-\Phi^i(\tau)}{\lambda}=\frac{\d\Phi^i
}{\d \ln\tau}\,.
\end{equation}
Thus a scaling solution is associated with an invariance of the
equations of motion for a rescaling of cosmic time and is therefore
associated with a conformal Killing vector on space-time and a
Killing vector on the scalar manifold.

Pseudo-supersymmetry comes into play when we check the geodesic
equation of motion
\begin{equation}
\nabla_{\dot{\Phi}}\dot{\Phi}_i=\dot{\Phi}^j\nabla_j\dot{\Phi}_i=\dot{\Phi}^j\Bigr\{\nabla_{(j}\dot{\Phi}_{i)}
+ \nabla_{[j}\dot{\Phi}_{i]}\Bigl\}\,,
\end{equation}
where we denote $\dot{\Phi}_i=G_{ik}\dot{\Phi}^k$. Now we have that
the symmetric part is zero if we parametrize the curve with
$t=\ln\tau$ since scaling makes $\dot{\Phi}$ a Killing vector. We
also have that $\nabla_{[j}\dot{\Phi}_{i]}=0$ since the pseudo-BPS
condition makes $\dot{\Phi}$ a curl-free  flow
$\dot{\Phi}_i=-f\partial_i W$.  To check that the curl is indeed
zero (when $f\neq 1$) one has to notice that in the parametrization
of the curve in terms of $t=\ln \tau$ the gauge is such that
$\dot{g}/g$ is constant and that $f\sim W^{-1}$. Since the curl is
also zero we notice that the curve is a geodesic with $\ln\tau$ as
affine parametrization\footnote{ One could wonder whether the
results works in two ways. Imagine that a scaling solution is a
geodesic. This then implies that $\nabla_{[j}\dot{\Phi}_{i]}=0$ and
therefore the flow is locally a gradient flow
$\dot{\Phi}_i=\partial_i \ln W \sim f\partial_i W$.}
\begin{equation}
\nabla_{\dot{\Phi}}\dot{\Phi}^i=0=\ddot{\Phi}^i +
\Gamma^i_{jk}\dot{\Phi}^j\dot{\Phi}^k\,.
\end{equation}

The link between scaling and geodesics was discovered by Karthauser
and Saffin in \cite{Karthauser:2006ix}, but no conditions on the
Lagrangian were given in \cite{Karthauser:2006ix} such that the
relation scaling-geodesic holds. An example of a scaling solution
that is not a geodesic was given by Sonner and Townsend in
\cite{Sonner:2006yn}.

A more intuitive understanding of the origin of the geodesic motion
for some scaling cosmologies comes from the on-shell substitution
$V=(3p-1)\,T$ in the Lagrangian to get a new Lagrangian describing
seemingly massless fields. Although this is rarely a consistent
procedure we believe that this is nonetheless related to the
existence of geodesic scaling solutions.
\subsubsection*{Single field}
For single-field models the potential must be exponential
$V=\Lambda\e^{\alpha\phi}$ in order to have scaling solutions. The
simplest pseudo-superpotential belonging to an exponential potential
is itself exponential
\begin{equation}\label{pseudosuperpot}
W=\pm\sqrt{\tfrac{8\Lambda}{3-\alpha^2}}\,\,\e^{\tfrac{\alpha\phi}{2}}\,.
\end{equation}
If we choose the plus sign the solution to the pseudo-BPS equation
is
\begin{equation}
\phi(\tau)=-\tfrac{2}{\alpha}\ln\tau+\tfrac{1}{\alpha}\ln
[\tfrac{6-2\alpha^2}{\alpha^4\Lambda}]\,,\qquad
g(\tau)\sim\tau^{\tfrac{1}{\alpha^2}}\,.
\end{equation}
The minus sign corresponds to the time reversed solution.
\subsubsection*{Multiple fields}
For a general multi-field model a scaling solution with power-law
scale factor $\tau^p$ obeys $V=(3p-1)T$ from which we derive the
{\bf on-shell} relation
\begin{equation}
G^{ij}\partial_iW\partial_jW=\frac{W^2}{4p}\quad\Rightarrow\quad
W=\pm\sqrt{\frac{8\,p\,V}{3p-1}}\,.
\end{equation}
In general the above expression for the superpotential
$W\sim\sqrt{V}$ does not hold off-shell, unless the potential is a
function of a specific kind:
\begin{equation}\label{scalingpotential}
\frac{1}{p}=\frac{G^{ij}\partial_i V\partial_j V}{V^2}\,.
\end{equation}
Scalar potentials that obey (\ref{scalingpotential}) with the extra
condition that $p\gtrless\tfrac{1}{3}\leftrightarrow V\gtrless 0$
allow for multi-field scaling solutions. For a given scalar
potential that obeys (\ref{scalingpotential}) there probably exist
many pseudo-superpotentials $W$ compatible with $V$ but if we make
the specific choice $W=\sqrt{8\,p\,V/(3p-1)}$ then all pseudo-BPS
solutions must be scaling and hence geodesic. As a consistency check
we substitute the first-order pseudo-BPS equations into the
right-hand-side of the following second-order equations of motion
\begin{equation} \label{secondorder}
\ddot{\Phi}^i + \Gamma_{jk}^i\dot{\Phi}^k\dot{\Phi}^j  =-
f^2G^{ij}\partial_j V - \Bigl[3\dot{(\ln g)}-\dot{(\ln
f)}\Bigr]\dot{\Phi}^i\,,
\end{equation}
and choose a gauge for which
\begin{equation}
\frac{\dot{f}}{f^2}=\frac{1}{4p}W\,,
\end{equation}
then we indeed find an affine geodesic motion since the right hand
side of (\ref{secondorder}) vanishes.

For some systems one first needs to perform a truncation in order to
find the above relation (\ref{scalingpotential}). A good example is
the multi-field potential appearing in Assisted Inflation
\cite{Liddle:1998jc}
\begin{equation}
V(\Phi^1,\ldots,\Phi^n)=\sum_i^n \Lambda_i\,\e^{\alpha_i\Phi^i}\,,
\qquad G_{ij}=\delta_{ij}\,.
\end{equation}
The scaling solution of this system was proven to be the same as the
single-exponential scaling \cite{Malik:1998gy}. The reason is that
one can perform an orthogonal transformation in field space such
that the form of the kinetic term is preserved but the scalar
potential is given by
\begin{equation}
V=\e^{\alpha\varphi}\,U(\Phi^1,\ldots,\Phi^{n-1})\,,\quad
\frac{1}{\alpha^2}=\sum_{i}\frac{1}{\alpha_i^2}\,.
\end{equation}
The scaling solution is such that $\Phi_1,\ldots,\Phi_{n-1}$ are
frozen in a stationary point of $U$ and therefore the system is
truncated to a single-field system that obeys
(\ref{scalingpotential}). The same was proven for Generalized
Assisted Inflation \cite{Copeland:1999cs} in reference
\cite{Hartong:2006rt}. The scaling solution in the original field
coordinates reads $\Phi^i=A^i\ln\tau+B^i$, which is clearly a
straight line and thus a geodesic.

The scaling solutions of \cite{Sonner:2006yn,Sonner:2007cp} were
constructed for an axion-dilaton system with an exponential
potential for the dilaton
\begin{equation}\label{actionI}
S=\int \sqrt{-g}\Bigl\{\mathcal{R}-\tfrac{1}{2} (\partial\phi)^2
-\tfrac{1}{2}\e^{\mu\phi}
(\partial\chi)^2-\Lambda\e^{\alpha\phi}\Bigr\}\,.
\end{equation}
Clearly this two-field system obeys (\ref{scalingpotential}) and
(one of) the pseudo-superpotential(s) is given by
(\ref{pseudosuperpot}). The pseudo-BPS scaling solution therefore
has constant axion and is effectively described by the dilaton in an
exponential potential. Note that this solution indeed describes a
geodesic on $\SL(2,\Real)/\SO(2)$ with $\ln \tau$ as affine
parameter. All examples of scaling solutions in the literature seem
to occur for exponential potentials, however by performing a
$\SL(2,\Real)$-transformation on the Lagrangian (\ref{actionI}) the
kinetic term is unchanged and the potential becomes a more
complicated function of the axion and the dilaton. The same scaling
solution then trivially still exists (and (\ref{scalingpotential})
still holds) but the axion is not constant in the new frame and
instead the solution follows a more complicated geodesic on
$\SL(2,\Real)/\SO(2)$.

However another scaling solution is given in \cite{Sonner:2006yn}
that is not geodesic and with varying axion in the frame of the
above action (\ref{actionI}). This is an illustration of the above,
since the solution is not geodesic we know that there does not
exists any other pseudo-superpotential for which the varying axion
solution is pseudo-BPS, consistent with what is shown in
\cite{Sonner:2007cp} for that particular solution.

\section{Geodesic curves and the Borel gauge}

For the last example of the previous section the pseudo-BPS scaling
solutions described geodesics on the symmetric space
$\SL(2,\Real)/\SO(2)$. In this section we consider a general class
of symmetric spaces of which $\SL(2,\Real)/\SO(2)$ is an example and
they are known as maximally non-compact cosets $U/K$. It seems that
for this class of spaces the geodesic equations of motion can be
solved easily. The symmetry of the geodesic equations is the
symmetry of the scalar coset $U/K$. In the case of maximal
supergravity the symmetry $U$ is a U-duality and is a maximal
non-compact real slice of a complex semisimple group. The isotropy
group $K$ is the maximal compact subgroup of $U$.

\subsection{A solution-generating technique}

In the \emph{Borel gauge} the scalar fields are divided into $r$
dilatons $\phi^I$ and $(n-r)$ axions $\chi^{\alpha}$, with $r$ the
rank of $U$ and $n$ the dimension of $U/K$ (see for instance
\cite{Andrianopoli:1996bq}). The dilatons are related to the
generators $H_I$ of the Cartan sub-algebra (CSA) and the axions to
the positive root generators $E_{\alpha}$ through the following
expression for the coset representative $L$ in the Borel gauge
\begin{equation}
L=\Pi_{\alpha}\text{exp}[\chi^{\alpha}E_{\alpha}]\Pi_{I}\text{exp}[-\tfrac{1}{2}\phi^IH_I]\,.
\end{equation}
In this language the geodesic equation is
\begin{align}
\ddot{\phi}^I + \Gamma^{I}_{JK}\dot{\phi}^J\dot{\phi}^K +
\Gamma^I_{\alpha J}\dot{\chi}^{\alpha}\dot{\phi}^J +
\Gamma^I_{\alpha \beta}\dot{\chi}^{\alpha}\dot{\chi}^{\beta}& =
0\,,\\
\ddot{\chi}^{\alpha} + \Gamma^{\alpha}_{JK}\dot{\phi}^J\dot{\phi}^K
+ \Gamma^{\alpha}_{\beta J}\dot{\chi}^{\beta}\dot{\phi}^J +
\Gamma^{\alpha}_{\beta\gamma}\dot{\chi}^{\beta}\dot{\chi}^{\gamma} &
= 0\,.
\end{align}
Since $\Gamma_{JK}^I=0$ and $\Gamma_{JK}^{\alpha}=0$ at points for
which $\chi^{\alpha}=0$ a trivial solution is given by
\begin{equation}
\phi^I=v^I\,y\,,\qquad \chi^{\alpha}=0\,.
\end{equation}
How many other solutions are there? A first thing we notice is that
every global $U$-transformation $\Phi\rightarrow \tilde{\Phi}$
brings us from one solution to another solution. Since $U$
generically mixes dilatons and axions we can construct solutions
with non-trivial axions in this way. We now prove that in this way
\emph{all} geodesics are obtained and this depends on the fact that
$U$ is maximally non-compact with $K$ the maximal compact subgroup
of $U$.

Consider an arbitrary geodesic curve $\Phi(t)$ on $U/K$. The point
$\Phi(0)$ can be mapped to the origin $L=\mathbbm{1}$ using a
$U$-transformation, since we can identify $\Phi(0)$ with an element
of $U$ and then we multiply the geodesic curve $\Phi(t)$ with
$\Phi(0)^{-1}$, generating a new geodesic curve $\Phi_2(t)=
\Phi(0)^{-1}\Phi(t)$ that goes through the origin. The origin is
invariant under $K$-rotations but the tangent space at the origin
transforms under the adjoint of $K$. One can prove that there always
exists an element $k \in K$, such that $\text{Adj}_k
\dot{\Phi}_2(0)\in \text{CSA}$ \cite{Knapp}. Therefore
$\dot{\chi}^{\alpha}_2=0$ and this solution must be a straight line.
So we started out with a general curve $\Phi(t)$ and proved that the
curve $\Phi_3(t)= k\Phi(0)^{-1}\Phi(t)$ is a straight line.

\subsection{An illustration from dimensional reduction}

The metric Ansatz for the dimensional reduction of
$(4+n)$-dimensional Einstein-gravity on the $n$-torus
($\mathbbm{T}^n$) is
\begin{equation}\label{metriek}
\d s_{4+n}^2=e^{2\alpha \varphi}\d s_4^2 + e^{2\beta
\varphi}\mathcal{M}_{ab}\d z^a \otimes \d z^b\,,
\end{equation}
where
\begin{equation}\label{alpha en beta}
\alpha^2=\frac{n}{4(n+2)}\,,\qquad \beta=-\frac{2\alpha}{n} \,.
\end{equation}
The matrix $\mathcal{M}$ is a positive-definite symmetric $n\times
n$ matrix with unit determinant, which depends on the 4-dimensional
coordinates, describing the moduli of $\mathbb{T}^n$. The modulus
$\varphi$ controls the overall volume and is named the breathing
mode or radion field. Notice that we already truncated the
Kaluza--Klein vectors in the Ansatz. The reduction of the
Einstein--Hilbert term gives
\begin{equation}
\mathcal{L}=\sqrt{-g}\{\mathcal{R}-\tfrac{1}{2}(\partial\varphi)^2+\tfrac{1}{4}\text{Tr}\partial
\mathcal{M} \partial \mathcal{M}^{-1}\}\,.
\end{equation}
The scalars parametrize $\Real\times \SL(n,\Real)/\SO(n)$ where
$\varphi$ belongs to the decoupled $\Real$-part and $\mathcal{M}$ is
the $\SL(n,\Real)/\SO(n)$ part.

If we take the four-dimensional part of space-time to be a flat
FLRW-space then that part of the metric will be power-law with
$p=1/3$ and the scalars follow a geodesic with $\ln \tau$ as an
affine parameter. According to the solution-generating technique,
the Ansatz for the scalars is
\begin{equation}
\varphi=v_0 \ln \tau + c_0\,,\quad \mathcal{M}=\Omega D
\Omega^T\,,\quad D=\text{diag}(\e^{-\vec{\beta}_a \cdot
\vec{\phi}})\,,
\end{equation}
with $\vec{\phi}=\vec{v}\ln\tau$ and $\vec{\beta}$ the weights of
$\SL(n,\Real)$ in the fundamental representation (see appendix
\ref{SL} for some explanations on the $\SL(n,\Real)/\SO(n)$-coset in
this representation). The diagonal matrix $D$ represents the
straight-line solution and $\Omega$ is an arbitrary
$\SL(n,\Real)$-matrix in the fundamental representation. Therefore
$\mathcal{M}=\Omega D\Omega^T$ is the most general coset matrix
describing a geodesic curve.

The Friedmann equation implies that the affine velocity is
restricted to be
\begin{equation}\label{fried}
v_0^2+||v||^2=\tfrac{4}{3},
\end{equation}
which is the only constraint coming from the 4-dimensional Einstein
equation. If we substitute this solution in (\ref{metriek}) and
define new coordinates $\vec{y}=\vec{z}\,\Omega$ we find
\begin{equation}\label{metriekii}
\d s_{4+n}^2=-\tau^{2\alpha\,v_0}\d\tau^2 +
\tau^{\tfrac{2}{3}+2\alpha\,v_0} \d \vec{x}_3^2 +
\sum_{a=1}^n\tau^{-\vec{\beta}_a\cdot\vec{v}+2\beta\,v_0}\d y_a^2\,.
\end{equation}
This is similar to what is called a Kasner solution in general
relativity (see for instance \cite{Kokarev:1995ri}). Kasner
solutions are a general class of time-dependent geometries that look
like
\begin{equation}
\d s^2=-\tau^{2p_0}\d \tau^2 + \sum_a \tau^{2p_a}\d x_a^2\,.
\end{equation}
Kasner solutions solve the Einstein equations in vacuum if the
following two conditions are satisfied
\begin{equation}
 p_0+1    =\sum_a p_a\,,\,\,\,\,\,(p_0+1)^2=\sum_a p_a^2\,.
\end{equation}
For the metric (\ref{metriekii}) these conditions are satisfied if
the lower-dimensional Friedmann equation is satisfied. For this
calculation one needs the properties of the weight-vectors
$\vec{\beta}_a$ (given in appendix \ref{SL}) and the relation
between $\alpha$ and $\beta$ (\ref{alpha en beta}). We therefore
conclude that the general spatially flat FLRW-solution lifts up to
the most general Kasner solution with $\SO(3)$-symmetry in $4+n$
dimensions.

\section{Discussion}

In this note we have studied multi-field scaling solutions using a
first-order formalism for scalar cosmologies \emph{a.k.a.}
pseudo-supersymmetry. We derived these first-order equations via a
Bogomol'nyi-like method that was known to work for domain wall
solutions as was first shown in
\cite{Bakas:1999ax,Skenderis:1999mm}\footnote{See also
\cite{Bakas:1999fa}.} and we showed that it trivially extends to
cosmological solutions. This first-order formalism allows a better
understanding of the geodesic motion that comes with a specific
class of scaling solutions. One of the main results of this note
is a proof that shows that all pseudo-BPS cosmologies that are
scaling solutions must be geodesic. This complements to the
discussion in \cite{Sonner:2007cp} where the first example of a
non-geodesic scaling cosmology was shown to be non-pseudo-BPS.
Moreover we gave constraints on multi-field Lagrangians for which
the pseudo-BPS cosmologies are geodesic scaling solutions.

Having illustrated the importance of geodesic motion in scalar
cosmology, we tackled the problem of solving the geodesic equations
in the second part of this note. We showed that the most general
geodesic curve can be written down for maximally non-compact coset
spaces $U/K$. These coset spaces appear in all maximal and some
less-extended supergravities \cite{Fre:2006eu}. We used a
solution-generating technique based on the symmetries of the coset.
We were able to prove that the most general solution is given by a
U-transformation on the ``straight line'', ($\phi^I(t)=v^I t,
\chi^{\alpha}=0$) in the Borel gauge.  We illustrated this technique
for the coset $\SL(n,\Real)/\SO(n)$. Since $\SL(n,\Real)/\SO(n)$ is
also the moduli space of the $n$-torus we applied it to find the
cosmological solutions of higher-dimensional gravity compactified on
a $n$-torus. This exercise nicely illustrates why the straight line
is the generating solution since, from a higher-dimensional point of
view, all solutions that correspond to the non-straight line
geodesics can be seen as coordinate transformations of the solutions
associated with the straight line. The oxidation of the straight
line solutions corresponds to the most general $\SO(3)$-invariant
Kasner solution of $(4+n)$-dimensional vacuum GR.

The same technique was used in \cite{Rosseel:2006fs} to find all
geodesic scaling cosmologies of the $\CSO$-gaugings in maximal
supergravity.

The solution-generating technique presented here should be
considered complementary to the ``compensator method'' developed by
Fr\'e et al in \cite{Fre:2003ep}. There the straight line also
serves as a generating solution but instead of rigid
$U$-transformations one uses local $K$ transformations that preserve
the solvable gauge to generate new non-trivial solutions. This
technique is a nice illustration of the integrability of the
second--order geodesic equations of motion \cite{Fre:2005bs}.

\section*{Acknowledgments}
We are grateful to Dennis Westra for useful discussions and comments
on the manuscript and to Jan Rosseel for many useful discussions.
This work is supported in part by the European Community's Human
Potential Programme under contract MRTN-CT-2004-005104 in which the
authors are associated to Utrecht University. The work of AP and TVR
is part of the research programme of the "Stichting voor
Fundamenteel Onderzoek der Materie" (FOM).

\appendix
\section{Curvatures}\label{curvature}
For the metric Ansatz (\ref{ansatz}) the Ricci tensor is given by
\begin{equation}
\mathcal{R}_{ab}= -\epsilon(\eta_{\epsilon})_{ab}
\Bigr\{\frac{\d}{\d y}[\frac{g \dot{g}}{f^2}] +
\frac{g\dot{g}\dot{f}}{f^3}+(D-3)\frac{\dot{g}^2}{f^2} \Bigl\}
\,,\qquad \mathcal{R}_{yy}=
(D-1)\Bigr\{-(\frac{\ddot{g}}{g})+\frac{\dot{g}\dot{f}}{gf}\Bigl\}\,.
\end{equation}

\section{The coset $\SL(N,\Real)/\SO(N)$} \label{SL}

Consider a general coset $U/K$. It is not difficult to construct a
coset representative using the Lie algebras $\mathfrak{U}$ and
$\mathfrak{K}$ of $U$ and $K$ respectively. Since $K$ is a subgroup
of $U$ we have the decomposition $\mathfrak{U}=\mathfrak{K}\oplus
\mathfrak{F}$, with $\mathfrak{F}$ the complement of $\mathfrak{K}$
in $\mathfrak{U}$. For a given representation of the algebra
$\mathfrak{U}$ we define a coset representative via
$L(y)=\text{exp}(y^i \bold{f_i})$ where the $\bold{f_i}$ form a
basis of $\mathfrak{F}$ in some representation of $\mathfrak{U}$.

To derive the metric we define a Lie algebra valued one-form from
the coset representative $L(y)$ via
\begin{equation}
L^{-1}\d L  \equiv E + \Omega\,,
\end{equation}
where $E$ takes values in $\mathfrak{F}$ and $\Omega$ in
$\mathfrak{K}$. We notice that $L^{-1}\d L$ is invariant under left
multiplication with a $y$-independent element $g\in U$. Multiplying
$L$ from the right with local elements $k \in K$ results in
\begin{equation}
 E \rightarrow k^{-1}\,E\,k\,, \qquad \Omega \rightarrow k^{-1}\,\Omega \,k + k^{-1}\d
 k\,.
\end{equation}
In supergravity the parameters $y^i$ are scalar fields that depend
on the space-time coordinates $y^i=\phi^i(x)$. The one-form
$L^{-1}\d L$ can be written out in terms of coset-coordinate
one-forms $\d \phi^i$ which themselves can be pulled back to
space-time coordinate one-forms $\d \phi^i=\partial_{\mu} \phi^i \d
x^{\mu}$. Now we can write
\begin{equation}
L^{-1}\d L= E_{\mu}\d x^{\mu}+\Omega_{\mu}\d x^{\mu}\,.
\end{equation}
Under the $\phi$-dependent $K$-transformations $k(\phi(x))$ we have
that $\Omega_{\mu} \rightarrow k^{-1}\Omega_{\mu}k+
k^{-1}\partial_{\mu}k$ and $E_{\mu}\rightarrow k^{-1}E_{\mu}k$. It
is clear that $E_{\mu}$ is covariant under local $K$-transformations
and $\Omega_{\mu}$ transforms like a connection. Using this
connection $\Omega_{\mu}$ we can make the following $K$-covariant
derivative on $L$ and $L^{-1}$
\begin{equation}
 D_{\mu}L=\partial_{\mu}L-L\Omega_{\mu}\,,\qquad D_{\mu}L^{-1}= \partial_{\mu}L^{-1}+\Omega_{\mu}L^{-1}\,.
\end{equation}
To find a kinetic term for the scalars we notice that the object
\begin{equation}
\text{Tr}[D_{\mu}LD^{\mu}L^{-1}]=-\text{Tr}[E_{\mu}E^{\mu}]\,,
\end{equation}
has all the right properties as it contains single derivatives on
the scalars, it is a space-time scalar, it is invariant under rigid
$U$ transformations and under local $K$-transformations. Thus,
\begin{equation}
e^{-1}\mathcal{L}_{\text{scalar}}=-\text{Tr}[E_{\mu}E^{\mu}]\equiv
-\tfrac{1}{2}g(\phi)_{ij}\partial_{\mu} \phi^i\partial^{\mu}
\phi^j\,.
\end{equation}

If  SO$(N)$ is the maximal compact subgroup of $U$ and we work in
the fundamental representation, then the Lie algebra of $\SO(N)$ is
the vector space of  antisymmetric matrices,
\begin{equation}\label{action}
E=\frac{L^{-1}\d L+ (L^{-1}\d L)^T}{2}\,,\qquad
\Omega=\frac{L^{-1}\d L-(L^{-1}\d L)^T}{2}\,,
\end{equation}
and a calculation shows that
\begin{equation}\label{actionII}
e^{-1}\mathcal{L}_{\text{scalar}}=-\text{Tr}[E^2]=+
\tfrac{1}{4}\text{Tr}[\partial\mathcal{M}\partial\mathcal{M}^{-1}]\,,
\end{equation}
where $\mathcal{M}$ is the $\SO(N)$-invariant matrix
$\mathcal{M}=LL^T$.

No we specify to $U=\SL(N,\Real)$. In general $\SL(N,\Real)$ has
rank $N-1$ and its maximal compact subgroup is $\SO(N)$. There will
therefore be $N-1$ dilaton fields $\phi^I$ and $N(N-1)/2$ axion
fields $\chi^{\alpha}$. The Cartan generators are given in terms of
the weights $\vec{\beta}$ of $\SL(N,\Real)$ in the fundamental
representation
\begin{equation}
(\vec{H})_{ij}=(\vec{\beta}_i)\delta_{ij}\,.
\end{equation}
The weights can be taken to obey the following algebra
\begin{equation}\label{SLWEIGHTS}
\sum_i\beta_{iI}=0\,,\quad
\sum_i\beta_{iI}\beta_{iJ}=2\delta_{IJ}\,,\quad
\vec{\beta}_i\cdot\vec{\beta}_j=2\delta_{ij}-\frac{2}{N}\,.
\end{equation}
The first of these identities holds in all bases since it follows
from the tracelessness of the $\SL$ generators. The second and third
identity can be seen as convenient normalizations of the generators.
The positive step operators $E_{ij}$ are all upper triangular and a
handy basis is that they have only one non-zero entry
$[E_{ij}]_{ij}=1$. The negative step operators are the transpose of
the positive. The $\SO(N)$ algebra is spanned by the following
combinations
\begin{equation}
\frac{1}{\sqrt{2}}(E_{\beta}-E_{-\beta})\,.
\end{equation}
The action will generically look complicated but when all axions are
set to zero $L$ is diagonal
$L=\text{diag}[\,\text{exp}(-\tfrac{1}{2}\vec{\beta}_i\cdot\vec{\phi})]$
and the action becomes
\begin{equation}
+\tfrac{1}{4}\text{Tr}\partial\mathcal{M}\partial\mathcal{M}^{-1}=-\tfrac{1}{4}(\sum_i\beta_{iJ}\beta_{iI})\partial
\phi^I\partial\phi^J=-\tfrac{1}{2}\delta_{IJ}\partial
\phi^I\partial\phi^J\,.
\end{equation}
This action describes $N-1$ dilatons that parametrize the flat
scalar manifold $\Real^{N-1}$.

\bibliography{pseudo}
\bibliographystyle{utphysmodb}

\end{document}